\documentclass[doublecol]{epl2}
\title{Coexistence of CDW with staggered superconductivity in
a ferromagnetic material}
\author{M. Georgiou \inst{1} \and G. Varelogiannis \inst{1} \and P. Thalmeier \inst{2}}
\institute{
  \inst{1} Department of Physics, National Technical University of Athens, GR-15780 Athens, Greece\\
  \inst{2} Max Planck Institute for Chemical Physics of Solids, 01187 Dresden, Germany}

\pacs{74.20.-z}{Theories and models of superconducting state}
\pacs{74.20.Rp}{Pairing symmetries (other than s-wave)}
\abstract{
In usual superconductivity (SC), the pairs have zero total momentum
irrespective of their symmetry. Staggered SC would involve, instead,
pairs with a finite commensurate total momentum,
but such exotic states have never been proven to be realized in
nature. Here we study for the first time the influence of
particle-hole asymmetry on the competition of staggered SC with
Charge Density Waves (CDW) in a ferromagnetic medium. We obtain
unprecedented situations in which CDW and staggered SC {\it
coexist}. We also obtain cases of a SC dome near the collapse
of a CDW state as well as cascades of transitions that exhibit
remarkable similarities with the pressure phase diagram in UGe$_2$
suggesting that SC in this material may be {\it
staggered} coexisting and competing with a CDW state.}

\begin{document}

\maketitle

The field of unconventional superconductivity (SC) is an
extraordinarilly rich source of challenging problems for fundamental
and applied physics. High-$T_c$ cuprates, heavy fermion materials,
borocarbides and organic SC are examples of unconventional SC. A
multitude of unconventional SC states have been proposed but only
few have been proven to be realized in real material systems.
Singlet or triplet unconventional SC states considered as realized
so far are characterized by a zero total momentum of the pairs
indicating in fact that the superfluid density is homogeneous in
momentum space. In the present Letter we suggest that the surprising
ferromagnetic (FM) SC state of UGe$_2$ is instead {\it staggered},
in which case the pairs have {\it a finite total momentum}. We also
prove that staggered SC may {\it coexist} with charge density wave
(CDW) states explaining features of the pressure phase diagram in
UGe$_2$ and opening a new perspective for the discussion of other
unconventional SC.

The unexpected discovery of SC well inside the {\it itinerant} FM
state of UGe$_2$ under pressure \cite{Saxena} and subsequently in
ZrZn$_2$ \cite{PfleidererNat} and URhGe \cite{Aoki} represent a
fascinating challenge for our understanding of SC
\cite{Machida,Abrikosov,Fomin,Karchev,Miyake,Walker,
Mineev,Lonzarich,Belitz}.
%because FM and SC are both carried by itinerant electrons belonging
%to the same Fermi surface (FS) sheet and their simultaneous presence
%would be apriori excluded.
The kind of unconventional SC involved as well as the complexity of
the pressure phase diagrams in UGe$_2$ remain a puzzle. In fact,
below T$_C$(0) = 52 K an almost fully polarized FM state is
observed. Applying hydrostatic pressure, FM is suddenly eliminated
at p$_{c2}$ = 1.5 GPa. Around p$_{c1}$ = 1.1 GPa, SC appears and at
the optimum p$^*$ = 1.2 GPa one finds T$_c(p^*)\leq$ 1 K. This is
well in the FM state where the Curie temperature and FM moment are
still significant with T$_C$(p$^*$) $\simeq$ 30 K, i.e. 60\% of the
original T$_C$(0). In addition to SC, presumably another phase is
present inside the FM state below T$^*(0)$ = 30 K, and T$^*(p)$ also
decreases with pressure until at p$^*$ it hits the optimum T$_c$(p)
of the SC dome \cite{Pfleiderer,Miyake}. The nature of the T$^{*}$
phase is not clear, one possibility that we adopt here is to
associate it with a charge density wave (CDW)
\cite{Saxena,Huxley,Miyake,Nishioka}, a scenario supported also by
LDA+U calculations \cite{Pickett}. This would explain the associated
experimentally observed heat-capacity anomalies \cite{Tateiwa} and
jumps in the magnetization on crossing the T$^*$ phase boundary
\cite{Pfleiderer}. Finally, recent NQR experiments suggest that the
$T^*$ phase survives even below the SC transition dividing the SC
dome into two parts: a low pressure part where SC and the $T^*$
order coexist and a high pressure part where there is no signature
of the $T^*$ order \cite{Flouquet,NQR}.

We explore in this Letter the {\it competition of staggered SC (i.e.
zone boundary SC) with CDW} in a strong ferromagnetic background, a
situation that has never been considered before. In fact, all
previous theoretical investigations of SC in UGe$_2$ considered zero
momentum (or zone center) SC. Note that the possibility of the
relevance of staggered SC states for some heavy fermion compounds
has first been considered in the past by D.L. Cox and coworkers
\cite{Cox} in the context of a Ginzburg-Landau approach. Moreover,
the SC pairing of spinless (or single-spin) fermions has been
introduced in \cite{Rudd} and discussed within a Ginzburg-Landau
theory. Here we start from a mean field BCS-type Hamiltonian
\begin{eqnarray}
\label{BCS}
H = \sum_{\bf{k}}\xi_{\bf{k}} c^{\dagger}_{\bf{k}}c_{\bf{k}}-
\sum_{\bf{k}}\bigl( W_{\bf{k}}
c^{\dagger}_{\bf{k}}c_{\bf{k}+\bf{Q}} + h.c.\bigr) \nonumber \\
-
\sum_{\bf k} \bigl( \Delta^{\bf 0}_{\bf k}
c^{\dagger}_{\bf{k}}c^{\dagger}_{-\bf{k}} + h.c.\bigr)
-
\sum_{\bf k} \bigl(\Delta^{\bf Q}_{\bf k}c^{\dagger}_{\bf{k}}
c^{\dagger}_{-\bf{k}-\bf{Q}}+ h.c.\bigr)
\label{BCS}
\end{eqnarray}
The first term describes a 2D tight binding FS whose nesting
properties with a wave vector \v Q=($\pi,\pi$) are controlled by the
ratio of n.n. (t$_1$) and n.n.n. (t$_2$) hopping matrix elements.
For t$_2$/t$_1 < 1$ the nesting with \v Q is removed completely and
the FS changes its shape. This is a schematic model for the
destruction of nesting of the LDA+U Fermi surface \cite{Pickett}
under pressure. There are several possibilities of competing SC and
CDW orders described by the gap functions W and $\Delta$ in Eq.(\ref{BCS}).
They are related to the effective interactions via gap equations like
Eqs.~(\ref{Gap1},\ref{Gap2}).
The effective  interactions V$^{SC}_{k,k'}$  and V$^{CDW}_{k,k'}$ of the
itinerant 5f-quasiparticles have a purely electronic origin. We may
have both unconventional SC with zero total pair momentum
$\Delta^{\bf 0} _{\bf k}$ and at finite pair momentum $\Delta^{\bf
Q}_{\bf k}$. The CDW gap function is denoted by W$_{\bf k}$ and like
$\Delta^{\bf 0} _{\bf k}$ or $\Delta^{\bf Q} _{\bf k}$ belongs to an
irreducible representation of the tetragonal D$_{4h}$ group (this is
also the approximate symmetry of UGe$_2$).

To treat both SC and CDW order parameters in a compact
manner we introduce a Nambu-type representation using the spinors
%
%\begin{eqnarray}
$\Psi^{\dagger}_{\bf{k}}=\bigl(c^{\dagger}_{\bf{k}},
c_{-\bf{k}},c^{\dagger}_{\bf{k}+\bf{Q}},c_{-\bf{k}-\bf{Q}}\bigr) $.
%\end{eqnarray}
%
Accordingly we use the tensor products
$\widehat{\rho}=\bigl(\widehat{\sigma}\otimes \widehat{I})~
\mbox{and}~ \widehat{\sigma}=\bigl(\widehat{I}\otimes
\widehat{\sigma})$ for the Nambu representation of the Hamiltonian
in Eq.~(\ref{BCS}). We assume that nesting in the fully FM polarized
band is responsible for the CDW transition associated with the $T^*$
line in UGe$_2$. Pressure reduces $T^*$ because it relaxes the
nesting conditions. To model this effect we write the electron
dispersion as a sum of particle-hole symmetric terms responsible for
nesting and particle-hole asymmetric terms that represent the
deviations from nesting: $\xi_{\bf {k}}=\gamma_{\bf {k}}+\delta_{\bf {k}}$
where $2\gamma_{\bf {k}}=\xi_{\bf {k}}-\xi_{\bf{k}+\bf{Q}}$ and
$2\delta_{\bf {k}}=\xi_{\bf {k}}+\xi_{\bf{k}+\bf{Q}}$. When $\delta_{\bf{
k}}=0$ there is particle-hole symmetry or perfect nesting with
wavevector $\bf{Q}$. Application of pressure adds a $\delta_{\bf {k}}$
term in the dispersion in addition to the  $\gamma_{\bf {k}}$ term
already present at zero pressure. We classify the SC and CDW order
parameters with respect to their behavior under inversion (I) ${\bf {
k}} \rightarrow {\bf {-k}}$, translation (t$_{\bf {Q}}$) ${\bf {k}}
\rightarrow {\bf k+Q}$ and time reversal (T) in the charge sector.
Instead of the latter we may also use complex conjugation (C) which
satisfies the equivalence relations C $\equiv$ -T ($\Delta^0_{\bf {
k}}$); C $\equiv$ IT ($\Delta^0_{\bf {k}}$) or C $\equiv$ t$_{\bf {Q}}$
(W$_{\bf {k}}$). These discrete transformations may then be used to
classify the possible groups of competing SC/CDW order parameters.
Obviously C is redundant for the three order parameters considered,
but we include it in the notation for clarity.

Because the spins are frozen, the  $\bf {q}=0$ SC pair states may only have
odd parity with $\Delta^{\bf {0}}_{-\bf {k}} = -\Delta^{\bf {0}}_{\bf {
k}}$. Under translation we have both signs $\Delta^{\bf {0}}_{\bf {k} + \bf {Q}}=
\pm\Delta^{\bf {0}}_{\bf {k}}$ and under C we get $(\Delta^{\bf {0}}_{\bf {k}}$)$^*$ =
- ($\Delta^{\bf {0}}_{\bf {k}}$)$^T$ = - $\Delta^{\bf {0}}_{\bf {k}}$. SC pair
states with finite momentum may in principle have both parities:
 $\Delta^{\bf {Q}}_{-\bf {k}} = \pm\Delta^{\bf {Q}}_{\bf {k}}$ because the
required antisymmetry may also come from the shift by a lattice
vector \textbf{R} with $\exp(i\bf Q\bf R)=-1$. On the other hand the t$_{\bf {
Q}}$ translation requires that always $\Delta^{\bf {Q}}_{\bf {k} + \bf {Q}}=
-\Delta^{\bf {Q}}_{\bf {k}}$ and under C we have  ($\Delta^{\bf {Q}}_{\bf {
k}}$)$^*$ = ($\Delta^{\bf {Q}}_{-\bf {k}}$)$^T$ = - $\Delta^{\bf {Q}}_{-\bf {
k}}$. These transformation properties allow {\it four possible SC
order parameters, two at zone center and two at zone boundary or
staggered SC}:
%
%\begin{eqnarray}
$ \Delta^{{\bf 0}I--}_{\bf k}, \hskip 0.3cm \Delta^{{\bf 0}I-+}_{\bf
k}, \hskip 0.3cm \Delta^{{\bf Q}R--}_{\bf k}, \hskip 0.3cm
\Delta^{{\bf Q}I+-}_{\bf k} $.
%\end{eqnarray}
%
Here the first index $\bf {0}$ or ${\bf {Q}}$ indicates the {\it total
momentum of the pair}, the second index $R$ or $I$ indicates whether
the order parameter is real or imaginary, the third index $\pm$
indicates parity under inversion I and the last index denotes gap
symmetry under t$_{\bf Q}$. As mentioned the index I(R) is redundant.

For the CDW order parameter both odd and even states under I and
t$_{\bf Q}$ are allowed so that $W_{-\bf {k}} = \pm W_{\bf {k}}$ and $ W_{\bf {
k}+\bf{Q}}=\pm W_{ \bf {k}}$ may hold. Since C $\equiv$ t$_ {\bf {Q}}$ for this
order parameter the redundant index R or I is associated with the
t$_{\bf {Q}}$-index $\pm$ respectively. As a result we have here again
four different possible (CDW) order parameters:
%
%\begin{eqnarray}
$ W^{R++}_{\bf {k}}, \hskip 0.3cm W^{I+-}_{\bf {k}}, \hskip 0.3cm
W^{R-+}_{\bf {k}}, \hskip 0.3cm W^{I--}_{\bf {k}} $,
%\end{eqnarray}
%
where the indices have the same meaning as the last three indices in
the SC order parameters. According to the above symmetry
classification there are {\it sixteen} possible pairs of such
competing SC/CDW states and only {\it eight of them concern
staggered SC} on which we are interested here. Within our formalism
we can calculate Green's functions and self-consistent gap equations
for each of these eight cases. As an example relevant for UGe$_2$
we report here for the
case of the competition of $W^{I+-}_{\bf {k}}$ with $\Delta^{{\bf
Q}R--}_{\bf {k}}$ where the Hamiltonian in spinor representation  with Pauli matrices
$\widehat{\sigma}_i $ and $\widehat{\rho}_i $ is
$ H =\sum_{\bf{k}}\Psi^{\dagger}_{\bf{k}} \widehat{\Xi}_{\bf {k}}
\Psi^{\dagger}_{\bf{k}} $ where
%
%\begin{eqnarray}
$ \widehat{\Xi}_{\bf k}=\gamma_{\bf {
k}}\widehat{\rho}_3\widehat{\sigma}_3 +\delta_{\bf {
k}}\widehat{\sigma}_3+\Delta^{{\bf {Q}}R--}_{\bf {k}}
\widehat{\rho}_2\widehat{\sigma}_2-W^{I+-}_{\bf {k}}\widehat{\rho}_2
$.
%\end{eqnarray}
%
We note that if instead of having the competition of $W_{\bf {
k}}^{I+-}\widehat{\rho}_2$ with $\Delta_{\bf{k}}^{{\bf{
Q}}R--}\widehat{\rho}_2\widehat{\sigma}_2$ as above, we had any of
the other pairs of competing order parameters, we would just have to
replace the corresponding SC and CDW terms in the above Hamiltonian.
The Green's functions that result would be modified accordingly. For
the $W_{\bf {k}}^{I+-}\widehat{\rho}_2$ and $\Delta_{\bf {k}}^{{\bf {
Q}}R--}\widehat{\rho}_2\widehat{\sigma}_2$ order parameters, the most
obvious realization in D$_{4h}$ symmetry is a d-wave CDW and p-wave
(finite momentum) SC order parameter given by
%
%\begin{eqnarray}
%\label{Gapmodel}
%$ \Delta^{\v Q}_{\v k}&=&\Delta^{\v Q}_0(\sin k_x + \sin k_y)$
$\Delta^{\bf {Q}}_{\bf {k}}=\Delta^{\bf {Q}}_0(\sin k_x + \sin k_y)$
%\qquad
(i.e. $E_u (1,1)$) and
%\nonumber\\
%$W_{\v k}&=&W_0(\cos k_x - \cos k_y)\qquad B_{1g}
$W_{\bf {k}}=W_0(\cos k_x - \cos k_y)$ (i.e. $ B_{1g}$).
%\end{eqnarray}
%
%For these competing gap functions the Green's function takes the form
%\begin{widetext}
%\begin{eqnarray}
%\widehat{G}({\bf k},i\omega_n)=\frac{-\bigl(
%i\omega_n+\widehat{\Xi}_{\bf k}\bigr) \bigl[A({\bf k},i\omega_n)
%-2\gamma_{\bf k}\delta_{\bf k}\widehat{\rho}_3+ 2\gamma_{\bf
%k}\Delta_{\bf k}\widehat{\rho}_1\widehat{\sigma}_1 +2\delta_{\bf
%k}W_{\bf k}\widehat{\rho}_2\widehat{\sigma}_3 +2\Delta_{\bf
%k}W_{\bf k}\widehat{\sigma}_2 \bigr]}{\bigl[
%\omega^2_n+E^2_+({\bf k}) \bigr]\bigl[\omega^2_n+E^2_-({\bf k})
%\bigr]}
%\end{eqnarray}
%\end{widetext}
%To simplify notations we have introduced the function
%$A({\bf k},i\omega_n)=\omega^2_n+\gamma^2_{\bf k}+ \delta^2_{\bf
%k}+\Delta^2_{\bf k} +W^2_{\bf k} $. The dispersion of quasiparticle
%branches $E_{\pm}$ is given by:
%
%\begin{eqnarray}
%$ E_{\pm}({\bf k})= \sqrt{\gamma^2_{\bf k}+W^2_{\bf k}}\pm
%\sqrt{\delta^2_{\bf k}+\Delta^2_{\bf k}} $
%\end{eqnarray}
%
From the Hamiltonians we obtain Green's functions and then
self-consistent gap equations for both order parameters which after
analytic summation over the Matsubara frequencies take the form of
the following system of coupled equations that are reported for the
first time here:
\begin{eqnarray}
\label{Gap1} \Delta_{\bf {k}}= \sum_{\bf {k'}}{ V^{SC}_{\bf k,k'}
\Delta_{\bf {k'}}\over 4 \sqrt{\delta^2_{\bf {k'}}+\Delta^2_{\bf {k'}}}}
\Bigl[ \tanh { E_{+}({\bf k'})\over 2T}- \tanh { E_{-}({\bf
k'})\over 2T}\Bigr]\\
\label{Gap2} W_{\bf k}=\sum_{\bf k'} {V^{CDW}_{\bf k,k'}W_{\bf k'}\over
4 \sqrt{\gamma^2_{\bf k'}+W^2_{\bf k'}}} \Bigl[ \tanh { E_{+}({\bf
k'})\over 2T}+ \tanh { E_{-}({\bf k'})\over 2T}\Bigr]\\
E_{\pm}({\bf k})= \sqrt{\gamma^2_{\bf k}+W^2_{\bf k}}\pm
\sqrt{\delta^2_{\bf k}+\Delta^2_{\bf k}}
\end{eqnarray}
Here the effective potentials  V$^{SC}_{k,k'} $, V$^{CDW}_{k,k'}$ are
separable for the asummed E$_u$ and B$_{1g}$ channels.
If solutions of these coupled SC/CDW gap equations exist they are unique.
For uniform order parameters assumed here they also have lower free energy
as compared to the normal state \cite{Machida81a,Machida81b}.
Eqs.~(\ref{Gap1},\ref{Gap2}) account for the following {\it
four} pairs of competing CDW and zone boundary states:
$\Delta^{RQ-}$ with $W^{R++}$, $\Delta^{RQ-}$ with $W^{I+-}$,
$\Delta^{IQ-}$ with $W^{R++}$ and $\Delta^{IQ-}$ with $W^{I+-}$.
There is a second system of coupled gap equations that describes the
competition of the {\it remaining four} pairs of SC and CDW gaps:
$\Delta^{RQ-}$ with $W^{R-+}$, $\Delta^{RQ-}$ with $W^{I--}$,
$\Delta^{IQ-}$ with $W^{R-+}$ and $\Delta^{IQ-}$ with $W^{I--}$:
%\begin{widetext}
\begin{eqnarray}
\label{gap3} \Delta_{\bf k}=\sum_{\bf k'} V_{\bf k k'}^{SC}
\Delta_{\bf k'}
%\nonumber\\
\biggl\{ {B({\bf k}) + \gamma_{\bf k'}^2 \over 4 E_{+}({\bf
k'})B({\bf k})}\tanh \biggl[ {E_{+} ({\bf k'})\over 2T} \biggr]
\nonumber\\
+ {B({\bf k}) - \gamma_{\bf k'}^2 \over 4 E_{-}({\bf k'})B({\bf
k})}\tanh \biggl[ {E_{-} ({\bf k'})\over 2T} \biggr]\biggr\}
\\
%\label{gap4}
%\end{eqnarray}
%\begin{eqnarray}
\label{gap4} W_{\bf k}=\sum_{\bf k'} V_{\bf k k'}^{CDW} W_{\bf k'}
%\nonumber\\
\biggl\{ {B({\bf k}) + \delta_{\bf k'}^2 \over 4 E_{+}({\bf
k'})B({\bf k})}\tanh \biggl[ {E_{+} ({\bf k'})\over 2T} \biggr]
\nonumber\\
+ {B({\bf k}) - \delta_{\bf k'}^2 \over 4 E_{-}({\bf k'})B({\bf
k})}\tanh \biggl[ {E_{-} ({\bf k'})\over 2T} \biggr]\biggr\}
%\label{gap4}
\end{eqnarray}
where
\begin{eqnarray}
E_{\pm}({\bf k})=\sqrt{ { \Delta_{\bf k}^2 W_{\bf k}^2\over
\gamma_{\bf k}^2 + W_{\bf k}^2}+ \biggl[ \sqrt{ \gamma_{\bf k}^2 +
W_{\bf k}^2}\pm \sqrt{{ B({\bf k})\over \gamma_{\bf k}^2 + W_{\bf
k}^2}}\biggr]^2}
\\
%\end{eqnarray}
%\begin{eqnarray}
B({\bf k})=\sqrt{\delta_{\bf k'}^2 \bigl( \gamma_{\bf k'}^2 + W_{\bf
k'}^2 \bigr) + \gamma_{\bf k'}^2\Delta_{\bf k'}^2} %\nonumber
\end{eqnarray}
%\end{widetext}

%%%%%%%%%%%%%%%%%%%%%%%%%%%%%%%%%%%%%%%%%%%%%%%%%%%%%%%%%%%%%%%%%%%%%%%%
\begin{figure}
\includegraphics[width=7.5cm,height=4.90cm,angle=0]{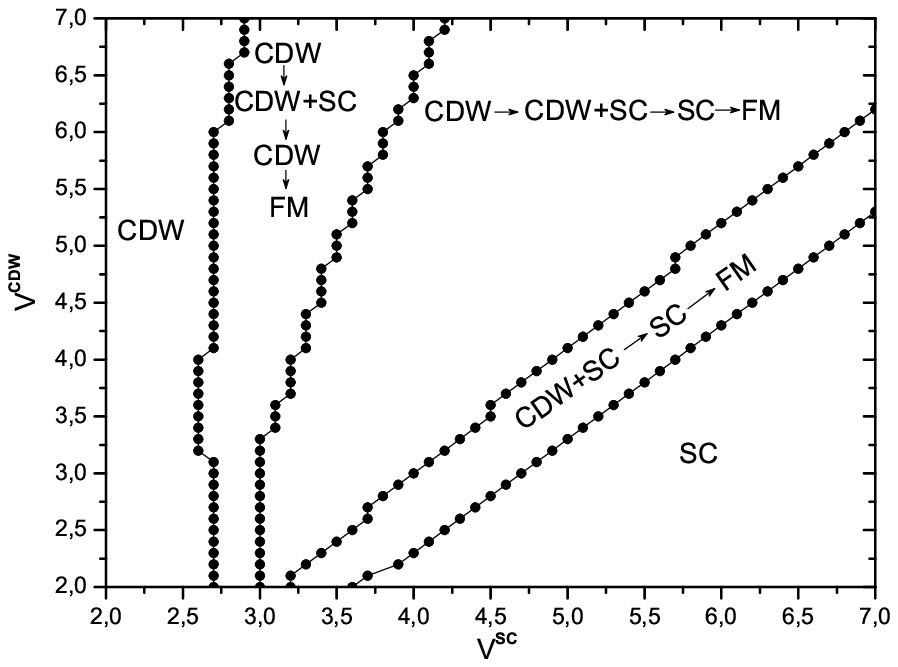}
\includegraphics[width=7.5cm,height=4.90cm,angle=0]{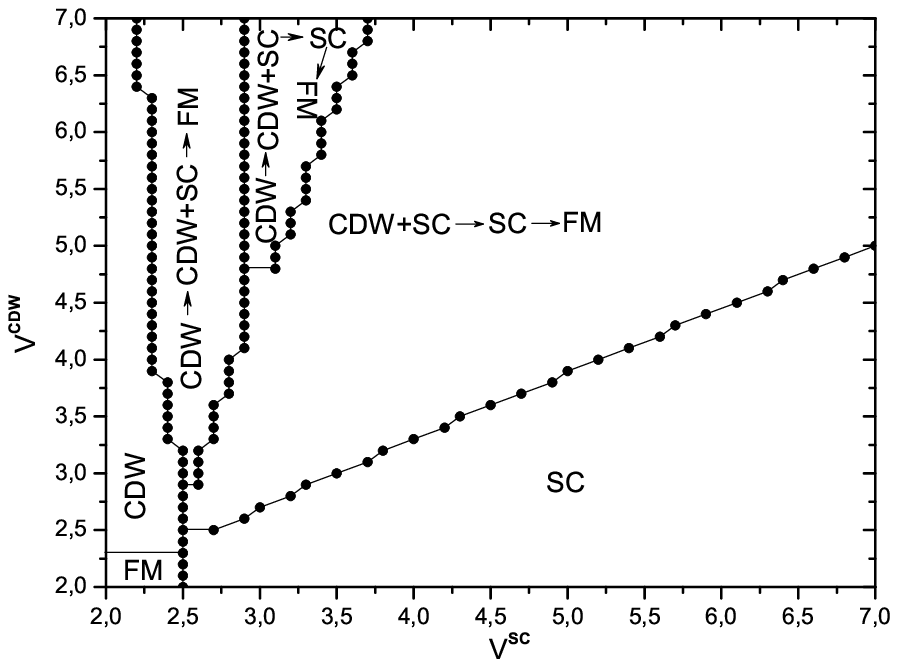}
\caption{Maps of the dependence of phase sequences on the effective interactions
$V^{CDW}$ and $V^{SC}$ for low temperature. Arrows indicate
the cascade of phases obtained when $t_2/t_1$ grows starting from zero. The black dots
separate regions of different phase {\it sequences} under growing $t_2/t_1$.
All phases coexist with ferromagnetism (FM). The phases
indicated as FM, are phases in which there is not any finite SC or
CDW order parameter and so only FM is present. Figure a) corresponds
to the competition of $\Delta^{RQ--}$ with $W^{I--}$. Figure b)
corresponds to the competition of $\Delta^{RQ--}$ with $W^{R-+}$.
The potentials are in units of $t_1$.} \label{fig:Contourphase}
\end{figure}
%%%%%%%%%%%%%%%%%%%%%%%%%%%%%%%%%%%%%%%%%%%%%%%%%%%%%%%%%%%%%%%%%%%%%%%%%
%
We have solved selfconsistently the systems of
Eqs.~(\ref{Gap1},\ref{Gap2}) and (\ref{gap3}, \ref{gap4}) for a 2D
tight-binding model on a square lattice. In that case, the
particle-hole symmetric term corresponds to nearest neighbor hoping
$\gamma_{\bf k}=t_1 (\cos k_x + \cos k_y)$ while particle-hole
asymmetry is introduced by the next-nearest neighbor hopping terms
$\delta_{\bf k}=t_2\cos k_x \cos k_y$. We have performed a large
number of self consistent calculations varying the pairing
potentials in the two channels producing eight maps (two of them
reported in figure 1) of {\it all possible transitions induced by
particle-hole asymmetry (i.e. by pressure)} in the low-$T$ region
for all the pairs of competing CDW and staggered SC order parameters
that are possible. To take into consideration the fact that the
different CDW and SC gap symmetries involved may correspond to
different momentum structures for the order parameters, we have
considered the separable potentials approximation that allows to
search for solutions of a specific momentum structure. Therefore,
the axes in Fig.1 are the amplitudes V$^{CDW}$ and V$^{SC}$ of the
pairing interactions and it is understood that a corresponding form
factor has been considered. We have investigated the coexistence of
order parameters in the whole (V$^{CDW}$, V$^{SC}$) plane from the
moderate coupling to the strong coupling regime since we have no
microscopic derivation for the effective pairing strengths. Arrows
in Fig. 1 indicate the cascade of phases observed when the ratio
$t_2/t_1$ grows starting from zero. Since we consider a spin
polarized background, all states reported also coexist with FM, and
the transitions to the FM state reported at high values of $t_2/t_1$
has the meaning of a transition to a state that is only
ferromagnetic with no CDW or SC order parameter present. We note in
figure 1a that in a large portion of the $V^{SC},V^{CDW}$ parameter
space there is at low temperature a transition from a CDW state to a
state in which CDW and SC coexist. The same transition is also
present over a portion of the parameter space in the case of figure
1b.
We note that the coexistence of zone-center SC with CDW has been reported in previous theoretical studies \cite{Machida84, Machida81a, Baeriswyl}.

%%%%%%%%%%%%%%%%%%%%%%%%%%%%%%%%%%%%%%%%%%%%%%%%%%%%%%%%%%%%%%%%%%%%%%%%
\begin{figure}
\includegraphics[width=7.5cm,height=4.90cm,angle=0]{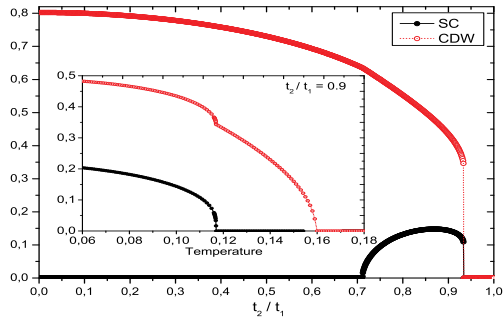}
\caption{(Color online): A characteristic example in which
deviations from nesting induced by $t_2/t_1$ lead to coexistence of
the $W^{R-+}$ CDW  (open circles in red) with the $\Delta^{RQ--}$
staggered SC order parameter (filed circles in black) corresponding
to $V^{CDW}=4$ and $V^{SC}=2.5$. We observe the cascade of
transitions from CDW to CDW+SC and then to FM as was already
reported in figure 1b in this region of of $V^{CDW}$ and $V^{SC}$
values. In the inset is shown the temperature dependence of both
order parameters when $t_2/t_1=0.9$ in which case the CDW and SC
orders coexist at low temperatures. All quantities are in units of
$t_1$.} \label{fig:Contourphase}
\end{figure}
%%%%%%%%%%%%%%%%%%%%%%%%%%%%%%%%%%%%%%%%%%%%%%%%%%%%%%%%%%%%%%%%%%%%%%%%%
%
We now look more closely to the situations in which CDW and
staggered SC may coexist. We report in figure 2a the behavior of the
CDW+SC state with $t_2/t_1$ in a characteristic example that
corresponds to $V^{CDW}=4$ and $V^{SC}=2.5$ in the competition of
$\Delta^{RQ--}$ with $W^{R-+}$ at low-T the mapping of which is
reported in figure 1b. We observe the cascade of transitions from
CDW to CDW+SC at $t_2/t_1\approx 0.72$ and finally to the FM state
for $t_2/t_1>0.93$ in agreement with the cascade reported in figure
1b for these couplings. It is remarkable in figure 2a that the
transition from CDW to CDW+SC is smooth as a function of $t_2/t_1$
in the low-T regime whereas the transition from the CDW+SC state to
the FM state (i.e. the state with no SC or CDW) is first order in
$t_2/t_1$. In figure 2b we show the corresponding transitions with
temperature when we take $t_2/t_1=0.9$. In this case, according to
figure 2a, we have indeed in low-T coexistence of CDW and SC. We
observe the counter-intuitive behavior that when SC appears as we
lower the temperature, the CDW gap grows instead of being reduced.

A particularly interesting cascade of transitions in relation to the
observations in UGe$_2$, is the one from CDW to CDW+SC to SC and
finally to FM. This cascade is observed over a large portion of the
parameter space of pairing potentials when the $\Delta^{RQ--}$ and
$W^{I--}$ compete (cf. fig. 1a) and over a smaller portion in the
case of competition of $\Delta^{RQ--}$ with $W^{R-+}$ (fig. 1b).
%%%%%%%%%%%%%%%%%%%%%%%%%%%%%%%%%%%%%%%%%%%%%%%%%%%%%%%%%%%%%%%%%%%%%%%%
\begin{figure}
\includegraphics[width=7.5cm,height=4.90cm,angle=0]{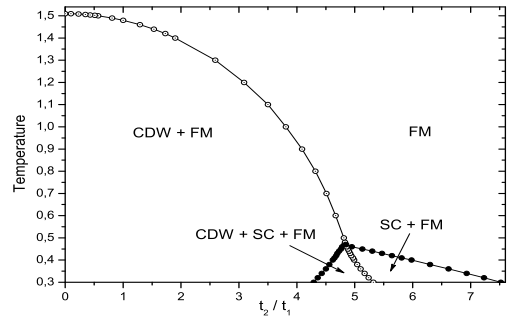}
\caption{Evolution with temperature of a cascade of $t_2/t_1$
induced transitions from CDW to CDW+SC to SC and then to only FM
obtained when $V^{CDW}=12.5$ and $V^{SC}=3.5$. The SC and CDW
order parameters considered are $\Delta^{RQ--}$ and $W^{R-+}$.
This phase diagram shows {\rm striking} similarities with features of the
pressure phase diagram of UGe$_2$ if one suppose that the $T^{x}$
phase corresponds to a $W^{R-+}$ CDW ordering and of course SC to
$\Delta^{RQ--}$. } \label{fig:Contourphase}
\end{figure}
%%%%%%%%%%%%%%%%%%%%%%%%%%%%%%%%%%%%%%%%%%%%%%%%%%%%%%%%%%%%%%%%%%%%%%%%%
%
A cascade of transitions in the low temperature regime that exhibits
amazing similarities with that observed in UGe$_2$ as a function of
pressure is shown in Fig.3. It corresponds to the competition of
$W^{R-+}$ CDW with the staggered SC order of the form
$\Delta^{RQ--}$ when $V^{SC}=3.5$ and $V^{CDW}=12.5$. The maximal
critical temperature of SC coincides with the crossing of the CDW
critical line. Moreover, the SC dome is divided into a part in which
SC and CDW coexist and a part in which only SC is present (with FM
of course). Quite remarkably the SC critical temperature is reduced
almost linearly at the higher values of $t_2/t_1$. The above results
are in surprising qualitative agreement with findings in UGe$_2$
\cite{Flouquet}. In particular, recent NQR results indicate that the
$T^*$ phase coexists indeed with SC over a portion of the SC dome
\cite{NQR} and our results are the first to provide a theoretical
picture for it.
%Naturally, the above picture could not explain why SC
%and FM are eliminated simultaneously in UGe$_2$ since we consider a
%spin frozen background and FM is built in the formalism.

Staggered SC states are relevant for magnetic SC
\cite{Cox,FuldeZwicknagl,Fenton} because they are similar to the
Fulde-Ferrel states \cite{FFLO} except that the modulation of the
superfluid density coincides with the characteristic wavevector of
the CDW. It appears, therefore, plausible to consider these states
in the analysis of SC in UGe$_2$ which is observed only in the FM
regime because for a staggered SC the FM background is necessary in
the same way as the magnetic field is necessary in order to obtain
the usual Fulde-Ferrel phase. Since pressure eliminates the FM state
it naturally eliminates simultaneously the staggered SC state as
well. Note finally that we obtain staggered SC states only over a
{\it limited dome} near the collapse of the CDW phase as in figures
2 and 3 {\it within a mean field approach without any fluctuations
involved}.

In conclusion, we have demonstrated the possibility to have {\it
coexistence of staggered SC with CDW} and cascades of transitions
induced by particle-hole asymmetry that reproduce the pressure phase
diagram observed in UGe$_2$ identifying the $T^*$ phase as a CDW
phase. Such exotic states may be relevant for other magnetic SC as
well.

\acknowledgments
We thank Jacques Flouquet and Modu Saxena for providing experimental
references. G.V. acknowledges visitor grants and hospitality from
the MPI CPfS at Dresden.

\end{document}